\documentclass{article}

\usepackage{arxiv}

\usepackage[utf8]{inputenc} % allow utf-8 input
\usepackage[T1]{fontenc}    % use 8-bit T1 fonts
\usepackage{hyperref}       % hyperlinks
\usepackage{url}            % simple URL typesetting
\usepackage{booktabs}       % professional-quality tables
\usepackage{amsfonts}       % blackboard math symbols
\usepackage{nicefrac}       % compact symbols for 1/2, etc.
\usepackage{microtype}      % microtypography
\usepackage{lipsum}		% Can be removed after putting your text content
\usepackage{graphicx}
\usepackage{natbib}
\usepackage{doi}

\usepackage{amsmath,amssymb,amsthm}
\usepackage[ruled,vlined,linesnumbered]{algorithm2e}
\usepackage{color}
\usepackage{multirow}
\usepackage{enumerate}
\usepackage{setspace}

\newcommand\mb[1]{\mathbf{#1}}
\newcommand\mc[1]{\mathcal{#1}}

\newcommand\bb[1]{\mathbb{#1}}
\newcommand\bs[1]{\boldsymbol{#1}}

\newcommand\wh[1]{\widehat{#1}}
\newcommand\whmb[1]{\widehat{\mathbf{#1}}}
\newcommand\whbs[1]{\widehat{\boldsymbol{#1}}}

\def\Var{\qopname\relax o{Var}}

\def\tr{\qopname\relax o{tr}}

\def\argmin{\qopname\relax m{argmin}}
\def\argmax{\qopname\relax m{argmax}}

\def\prox{\mathrm{prox}}

\newtheorem{assumption}{Assumption}
\newtheorem{theorem}{Theorem}

\newtheorem{corollary}{Corollary}
\newtheorem{remark}{Remark}
\newtheorem{definition}{Definition}

\title{CEDAR: Communication Efficient Distributed Analysis for Regressions}

\author{ Changgee Chang\thanks{Corresponding authors}\\
	Department of Biostatistics, Epidemiology, and Informatics\\
	University of Pennsylvania\\
	Philadelphia, PA 19104 \\
	\texttt{changgee@pennmedicine.upenn.edu} \\
	\And
	Zhiqi Bu \\
	Department of Biostatistics, Epidemiology, and Informatics\\
	University of Pennsylvania\\
	Philadelphia, PA 19104 \\
	\texttt{zbu@sas.upenn.edu} \\
	\And
	Qi Long\footnotemark[1] \\
	Department of Biostatistics, Epidemiology, and Informatics\\
	University of Pennsylvania\\
	Philadelphia, PA 19104 \\
	\texttt{qlong@upenn.edu} \\
}

\renewcommand{\shorttitle}{CEDAR}

\hypersetup{
pdftitle={A template for the arxiv style},
pdfsubject={q-bio.NC, q-bio.QM},
pdfauthor={Changgee Chang, Zhiqi Bu, Qi Long},
pdfkeywords={First keyword, Second keyword, More},
}

\begin{document}
\maketitle

\begin{abstract}
	Electronic health records (EHRs) offer great promises for advancing precision medicine and, at the same time, present significant analytical challenges. Particularly, it is often the case that patient-level data in EHRs cannot be shared across institutions (data sources) due to government regulations and/or institutional policies. As a result, there are growing interests about distributed learning over multiple EHRs databases without sharing patient-level data. To tackle such challenges, we propose a novel communication efficient method that aggregates the local optimal estimates, by turning the problem into a missing data problem. In addition, we propose incorporating posterior samples of remote sites, which can provide partial information on the missing quantities and improve efficiency of parameter estimates while having the differential privacy property and thus reducing the risk of information leaking. The proposed approach, without sharing the raw patient level data, allows for proper statistical inference and can accommodate sparse regressions. We provide theoretical investigation for the asymptotic properties of the proposed method for statistical inference as well as differential privacy, and evaluate its performance in simulations and real data analyses in comparison with several recently developed methods.
\end{abstract}

% keywords can be removed

\section{Introduction}
\label{sec:intro}

Recent advances in electronic health records (EHRs) databases have enabled us to collect massive EHRs data for healthcare delivery. Such rich, yet complex, data offer great promises for advancing precision medicine, but at the same time, bring some significant analytical challenges.
Suppose every hospital collects data from the admitted patients who suffered a stroke, and a hospital wishes to analyze the data to find out the risk factors that influence a certain outcome such as the arrival time to CT.
Unfortunately, the number of patients in the hospital can be small and the analysis results can be unreliable, even if the total number of stroke patients in the area can be collectively large.
A specious solution would be to borrow the de-identified patient data from other hospitals and incoporate them into the analysis to generate more reliable results.
However, this practice is often not feasible as the patient-level data, even after de-identification, cannot be shared across institutions due to the privacy of patients health.
One approach to address this challenge is the distributed health data networks (DHDNs, \citet{Maro2009}), through which the information of the data can be shared between participants but sharing the individual patient level data is not allowed.
One example of DHDNs is pSCANNER \citep{Ohno2014} which includes 13 data sites covering 37 million patients and has developed a suit of software tools for privacy-preserving distributed data analysis.

There are two important challenges in the distributed statistical analysis.
First, as mentioned, the individual patient level data cannot be exchanged to protect the privacy.
These include some of the summary statistics, which can potentially reveal the complete (or a subset of ) patients data.
Even providing the gradients of the loss function evaluated at multiple points may result in information leakage.
Second, the algorithm must be communication efficient, where we define the communication efficiency in terms of the number of communications between the central site and the remote sites (external data sources) to complete the entire analysis.
Requiring one-shot or only a few number of communications between sites is time and cost effective, given that the communications are typically manual.
Unfortunately, many of existing distributed learning algorithms require a large number of communications between sites, which can be feasible only if an automated communication system is available.
These challenges raise two important research questions; what kind of information could be shared from the external data sources while ensuring communication efficiency and privacy protection, and how the information can be incorporated into the analysis effectively.

Similar questions have been studied in the literature \citep{Zhang2013Information,Kleiner2014,Shamir2014,Mackey2015,Hector2021,Hector2020Doubly}.
Many of them focus on parallel computing involving iterative communications until convergence \citep{Maclaurin2015,Scaman2018}. Although some works address the privacy issue \citep{Abadi2016,Imtiaz2018}, those are typically neither communication efficient nor amenable to statistical inference.
The divide and conquer approaches, on the other hand, are communication efficient and often comes with a framework for statistical inference \citep{Lin2010,Zhang2013,He2016,Lee2017,Battey2018,Tang2020}, but they do not address the data privacy issue carefully.
This type of work also includes the recently proposed gradient enhanced loss function based approaches \citep{Wang2017,Jordan2019,Fan2021}.
In addition, vast majority of the existing distributed learning algorithms are developed for prediction tasks and do not allow for proper statistical inference such as hypothesis testing and inference on treatment effects.
Even for the linear regression model which is widely used in analysis of EHR data \citep{Shortreed2019}, there has been very limited work on developing privacy-preserving and communication-efficient methods.

In this paper, we seek to address this significant gap in the literature focusing on the linear regression model.
There are only a few existing methods that are communication efficient and amenable to statistical inference.
The simplest would be the average mixture (AVGM) of the local estimates or test statistics \citep{Zhang2013,Battey2018}, but these approaches are not amenable to sparse regressions such as lasso regression and are inflexible in that the testing hypotheses need to be specified before the communication occurs.
An alternative could be the global ordinary least square (OLS) estimator which can be constructed by transferring the local sufficient statistics to the central site.
However, even the sufficient statistic of linear regression may disclose patients' private data.
More recently, \citet{Jordan2019} proposed a communication efficient surrogate likelihood (CSL) approach, which only requires transferring the gradients of the local loss functions evaluated at an initial point.
One significant limitation of both AVGM and CSL is their poor finite sample performance, particularly when the site sample size is relatively small.

Our proposed method overcomes all the aforementioned limitations.
Our method aggregates the MLEs from the remote sites in a very clever way that has not been introduced before.
%Note that the sampling distribution of a local MLE depends not only on the model parameters but also on the local Gramian matrix, which cannot be shared.
One novelty of our work is to view the problem of distributed learning as a missing data problem, so that any solution to a missing data problem can become a potential solution to distributed analysis.
Another contribution of this work is to propose incorporating the posterior samples from the remote sites.
As noted, if all sufficient statistics for the external data were accessible, the global OLS estimator can be constructed.
The remote posterior samples can be seen as a natural random perturbation of the sufficient statistics and can supply partial (noisy) information about the sufficient statistic while mitigating the patients privacy issues.
\citet{Dimitrakakis2017} have shown that the posterior samples possess the differential privacy (DP) properties \citep{Dwork2014}, but how to incorporate the posterior samples into the analysis has not been answered in the literature.
In this work, we introduce an elegant and efficient algorithm that meets the need, which is called the Communication Efficient Distributed Analysis of Regressions (CEDAR).

The remainder of the article is organized as follows.
In Section \ref{sec:background}, we set up the problem and review three existing approaches.
In Section \ref{sec:CEDAR}, we propose our method and derive the algorithm.
In Section \ref{sec:theory}, we show the theoretical properties of the proposed method.
Performance is compared through comprehensive simulation studies in Section \ref{sec:simulation} and real data analyses in Section \ref{sec:data}, and  the article concludes with discussions in Section \ref{sec:discussion}.
The supplementary material includes all proofs for theoretical claims.

\section{Background}
\label{sec:background}

\subsection{Setup}
Suppose there are $M$ sites and each site has the data $(\mb y_m,X_m)$ of sample size $n$ and feature size $p$ for $m=1,\dots,M$. Here, $\mb y_m$ is the $n \times 1$ response vector and $X_m$ is the $n \times p$ design matrix. We assume a linear model with Gaussian errors.
\begin{align} \label{eqn:model}
\mb y_m = X_m \bs\beta_0 + \mb e_m,
\end{align}
where $\bs\beta_0$ is the $p \times 1$ true regression coefficient vector and $\mb e_m \sim \mc N(\mb 0, \sigma_0^2I)$ is the $n \times 1$ error vector.
Let $\mb x_i^m$ denote the $i$th row vector of $X_m$.
Although CEDAR does not need the normality for $\mb x_i^m$, we assume for now that $\mb x_i^m \sim \mc N(\mb 0,\Sigma_0)$ to help understanding of the derivation of the CEDAR algorithm which will be discussed in Section \ref{sec:CEDAR}.
This assumption will be relaxed in the discussion of theoretical properties of CEDAR in Section \ref{sec:theory}.
We assume the first site (site 1) is the central site, where the analysis is performed, and we have the full access to the local data of the central site.
We do not have access to the raw data of the remote sites but have limited access to some partial information which is method specific.
Let $N = nM$ be the total sample size.

\subsection{Related works}

In estimating the regression coefficients, the simplest approach would be the average mixture (AVGM) model \citep{Zhang2013}, which takes the average of the local OLS estimates.
\begin{align} \label{eqn:AVGM}
    \whbs\beta^A = \frac{1}{M} \sum_{m=1}^M \whbs\beta_m,
\end{align}
where $\whbs\beta_m = S_m^{-1} X_m^T \mb y_m$ with $S_m = X_m^TX_m$.
This is communication efficient because only one communication is required and it achieves the optimal convergence rate ($\sqrt{N}$-consistent). However, it is not efficient if $n \nrightarrow \infty$ since
\begin{align} \label{eqn:AVGMVar}
    \Var \left( \whbs\beta^A \right) = \frac{\sigma_0^2 \Sigma_0^{-1}}{M(n-p-1)}, \qquad n>p+1,
\end{align}
which explains its poor performance when the site sample size $n$ is small.

Another drawback of the AVGM approach is that statistical inference and sparse estimation are not straightforward.
\citet{Lee2017} simply applies the hard- and soft-thresholding on the average of the debiased lasso estimates \citep{zhang2014confidence,vandegeer2014,Javanmard2014}.
\citet{Battey2018} proposes inference in the spirit of AVGM.
For the Wald test, for example, it suggests
\begin{align} \label{eqn:AVGMWald}
\wh W^A = \frac{1}{\sqrt{M}} \sum_{m=1}^M \wh W_m \approx \mc N(0,1), 
\end{align}
where $\wh W_m$ are the standard Wald test statistics for the $m$-th dataset.
However, the hypothesis needs to be fixed before communication, and every additional test requires extra communications.

Alternatively, the sufficient statistic $(S_m, X_m^T\mb y_m, \mb y_m^T\mb y_m)$ can be transferred from all remote sites to the central site, and then the global OLS estimator can be constructed as follows.
\begin{align} \label{eqn:OPT}
    \whbs\beta^O = \left( \sum_{m=1}^M S_m \right)^{-1} \sum_{m=1}^M X_m^T\mb y_m,
\end{align}
which is asymptotically efficient regardless of the site sample size $n$ since
\begin{align} \label{eqn:OPTVar}
    \Var \left( \whbs\beta^O \right) = \frac{\sigma_0^2 \Sigma_0^{-1}}{N-p-1}, \qquad N>p+1.
\end{align}
This method is communication efficient and the inference is straightforward because the asymptotic variance can be estimated.
However, even the summary statistic $S_m$ can potentially reveal complete data for (a subset of) patients. Therefore, this method is not considered privacy preserving. 

Recently, \citet{Wang2017,Jordan2019} propose using, in addition to the full local loss function at the central site, the gradients of the remote loss functions evaluated at a certain initial point $\overline{\bs\beta}$, which is supposed to be close to $\bs\beta_0$.
They propose a new loss function called the communication-efficient surrogate likelihood (CSL), which is defined as
\begin{align} \label{eqn:CSL}
\widetilde{\mc L}(\bs\beta) = \mc L_1(\bs\beta) - ( \nabla \mc L_1 (\overline{\bs\beta}) - \nabla \mc L(\overline{\bs\beta}) )^T \bs\beta,
\end{align}
where $\mc L_m(\bs\beta) = \frac{1}{n} \sum_{i=1}^{n} \mc L(\bs\beta;\mb x_i^m,y_i^m)$ is the local average loss (negative loglikelihood) function at site $m$ and $\mc L(\bs\beta) = \frac{1}{M} \sum_{m=1}^M \mc L_m(\bs\beta)$ is the global average loss function, and the CSL estimator is defined as its minimizer; 
$\whbs\beta^C = \argmin_{\bs\beta} \widetilde{\mc L}(\bs\beta)$.

Note that the CSL estimator requires one round-trip communication in order to calculate $\nabla \mc L(\overline{\bs\beta})$.
The key idea behind \eqref{eqn:CSL} is to match the gradients of $\widetilde{\mc L}$ and $\mc L$ evaluated at $\overline{\bs\beta}$, i.e., $\nabla \widetilde{\mc L} (\overline{\bs\beta}) = \nabla \mc L(\overline{\bs\beta})$, hoping that the minimizers of the two functions become close to each other if $\overline{\bs\beta}$ is close to $\bs\beta_0$.
The solution achieves the optimal convergence rate if $\overline{\bs\beta}$ converges fast enough to $\bs\beta_0$ \citep{Jordan2019}.
However, we observe that the finite sample performance deteriorates if the initial estimator $\overline{\bs\beta}$ is not so accurate.
Note that $\overline{\bs\beta}$ is typically chosen to be the local solution of the central site, which is unstable if the site sample size $n$ is small.

\section{Methods}
\label{sec:CEDAR}

We propose our method CEDAR.
We first present how to aggregate the information from the remote MLEs, and then we propose how we can improve further by incorporating the remote posterior samples.

\subsection{Aggregating remote MLEs}
\label{sec:CEDAR0}

Suppose we have the local MLE $\{(\whbs\beta_1, \wh\sigma_1^2)\}$ and the remote MLEs $\{(\whbs\beta_m, \wh\sigma_m^2)\}_{2 \le m \le M}$ transmitted from remote databases. 
Note that the sampling distribution of the MLEs is given by
\begin{align} \label{eqn:sampling}
\wh\sigma_m^2 \sim \frac{\sigma_0^2}{n} \chi_{n-p}^2, \qquad \whbs\beta_m|S_m \sim \mc N(\bs\beta_0, \sigma_0^2 S_m^{-1}), \qquad S_m \sim \mc W(\Sigma_0,n).
\end{align}
Here, $\chi_\nu^2$ denotes the chi-square distribution with degrees of freedom $\nu$, and $\mc W$ stands for the Wishart distribution.
Note that we have assumed $\mb x_i^m \sim \mc N(\mb 0,\Sigma_0)$, in addition to the model assumption \eqref{eqn:model}, for the purpose of deriving the algorithm.
However, it is not a necessary condition for the theoretical properties of CEDAR, as it will be relaxed in Section \ref{sec:theory}.
Furthermore, the distribution of $\mb x_i^m$ may not be the same for all datasets.
Particularly, the covariance matrix of $\mb x_i^m$ may be heterogeneous across datasets.

The sampling distribution \eqref{eqn:sampling} yields the loglikelihood of $(\bs\beta,\sigma^2,\Sigma)$ as follows.
\begin{align*}
l_0(\bs\beta,\sigma^2,\Sigma) = & - \frac{N}{2} \log \sigma^2 - \frac{n}{2\sigma^2}\sum_m \wh\sigma_m^2 - \frac{N}{2} \log |\Sigma| - \frac{1}{2\sigma^2} \sum_m (\whbs\beta_m - \bs\beta)^T S_m (\whbs\beta_m - \bs\beta)\\
&+ \frac{n-p}{2} \sum_m \log |S_m| - \frac{1}{2} \tr \Sigma^{-1} \sum_m S_m.
\end{align*}
Note that this can serve as the density of the confidence distribution \citep{Xie2013,Schweder2016} for $(\bs\beta,\sigma^2,\Sigma)$ with additional random quantities $S_m$ for $m>1$.
If all $S_m$ were observed, it would be straightforward to find the mode (MLE) of $(\bs\beta,\sigma^2,\Sigma)$.
Since $S_m$ are not observed for $m>1$
($S_1$ is observed at the central site.),
our task is to find the MLE of $(\bs\beta,\sigma^2,\Sigma)$ in the presence of missing values $S_2,\dots,S_M$.
Any solution to missing value problems can potentially be used, and we take the strategy of marginalizing out those missing values and propose the MLE
\begin{align} \label{eqn:MLE}
(\whbs\beta,\wh\sigma^2,\wh\Sigma) = \argmax_{\bs\beta,\sigma^2,\Sigma} l(\bs\beta,\sigma^2,\Sigma),
\end{align}
where $l(\bs\beta,\sigma^2,\Sigma) = \log \int \cdots \int e^{l_0(\bs\beta,\sigma^2,\Sigma)} dS_2 \cdots dS_M$ is the loglikelihood.

We use the EM algorithm \citep{Dempster1977} to find the solution of \eqref{eqn:MLE}.
The E-step of EM algorithm requires to calculate the expected value of $l_0(\bs\beta,\sigma^2,\Sigma)$ with respect to the self conditional distribution given the model parameters.
Note that, conditioning on $(\bs\beta,\sigma^2,\Sigma)$, $\{S_m\}_{m>1}$ are independent and have the Wishart distribution as follows.
\begin{align*}
S_m | \bs\beta,\sigma^2,\Sigma \sim \mc W ((\Sigma^{-1} + \mb a_m \mb a_m^T)^{-1} ,n+1),
\end{align*}
where $\mb a_m = (\whbs\beta_m - \bs\beta) / \sigma$.
Therefore, the conditional expectation is given by
\begin{align*}
\bb E (S_m | \bs\beta,\sigma^2,\Sigma) = (n+1) \left(\Sigma - \frac{\Sigma \mb a_m \mb a_m^T \Sigma}{1 + \mb a_m^T \Sigma \mb a_m} \right).
\end{align*}
The M-step of EM algorithm optimizes the expected loglikelihood with respect to $(\bs\beta,\sigma^2,\Sigma)$, which yields the CEDAR algorithm described in Algorithm \ref{alg:CEDAR} with $K=0$ and $A_m = \mb a_m$. Here, $K$ refers to the number of posterior samples that can be additionally incorporated, which is discussed in the following section.

As shown in Section \ref{sec:simulation}, CEDAR with $K=0$ already works better than any existing method, particularly when the site sample size $n$ is relatively small.
However, as $p$ increases, the performance of CEDAR resembles that of AVGM, perhaps because it becomes difficult to accurately impute the $p \times p$ missing matrix $S_m$.
As a remedy, we propose incorporating remote posterior samples.

\begin{algorithm}
\caption{CEDAR Algorithm}
   \label{alg:CEDAR}

    {\bfseries At Site $m>1$:}\\
    Compute the MLE $(\whbs\beta_m,\wh\sigma_m^2)$ and the posterior samples $B_m$\;
    Send $(\whbs\beta_m,\wh\sigma_m^2,B_m)$ to the central site (Site 1) \tcp*{The only communication.}
    
    {\bfseries At Site $1$:}\\
    Find $(\whbs\beta_1,\wh\sigma_1^2,\wh S_1)$\;
    Receive $(\whbs\beta_m,\wh\sigma_m^2,B_m)$ from Sites $m=2,\dots,M$\;
    Initialize $\whbs\beta \leftarrow \frac{1}{M} \sum_m \whbs\beta_m$; $\wh\Sigma \leftarrow \frac{1}{n} \wh S_1$; $\wh\sigma^2 \leftarrow \frac{1}{N} \sum_m ((\whbs\beta_m - \whbs\beta)^T \wh S_1 (\whbs\beta_m - \whbs\beta) + n \wh\sigma_m^2)$\;
    
    \Repeat{convergence}{
        \tcc{Computed at the central site with no communication.}
        
        \For{$m\gets2$ \KwTo $M$}{
            $\whmb a_m \leftarrow (\whbs\beta_m - \whbs\beta) / \wh\sigma$;
            
            $\wh S_m \leftarrow (n+K+1) (\wh\Sigma^{-1} + \wh A_m \wh A_m^T)^{-1}$ where $\wh A_m = \begin{bmatrix} \whmb a_m & B_m/\sqrt{\psi} \end{bmatrix}$;
        }
        $\whbs\beta \leftarrow \left( \sum_m \wh S_m \right)^{-1} \sum_m \wh S_m \whbs\beta_m$;
        
        $\wh\sigma^2 \leftarrow \frac{1}{N} \sum_m ((\whbs\beta_m - \whbs\beta)^T \wh S_m (\whbs\beta_m - \whbs\beta) + n \wh\sigma_m^2)$;
        
        $\wh\Sigma \leftarrow \frac{1}{N} \sum_m \wh S_m$;
    }
    {\bfseries Output:} ($\whbs\beta$, $\wh\sigma^2$, $\wh\Sigma$, $\wh S_1$, \dots, $\wh S_M$);

\end{algorithm}

\subsection{Incorporating remote posterior samples}
\label{sec:CEDARK}

Suppose, in addition to the remote MLEs, we collect $K$ remote posterior samples $\{(\widetilde{\bs\beta}_{mk},\widetilde{\sigma}_{mk}^2)\}_{1\le k \le K}$ from each remote site $m=2,\dots,M$.
Each remote site generates the posterior samples from the scaled posterior density
$\pi_\psi(\bs\beta,\sigma^2|X_m, \mb y_m) \propto \pi(\mb y_m | \bs\beta,\sigma^2,X_m)^{1/\psi} \pi(\bs\beta,\sigma^2)$ where the prior $\pi(\bs\beta,\sigma^2) \propto (\sigma^2)^{-(p/2+1)}$ is used to ensure the scaled posterior to be proper.
These results in the posterior samples being generated under the following sampling distributions.
\begin{align*}
\widetilde{\sigma}_{mk}^2 \sim \mc I \mc G \left( \frac{n}{2\psi}, \frac{n \wh\sigma_m^2}{2\psi} \right), \qquad \widetilde{\bs\beta}_{mk}|\widetilde{\sigma}_{mk}^2 \sim \mc N(\whbs\beta_m, \psi \widetilde{\sigma}_{mk}^2 S_m^{-1}).
\end{align*}
Here, $\mc I \mc G$ stands for the inverse gamma distribution. These posterior samples are sent to the central site.

With these remote posterior samples combined at the central site, the (unmarginalized) loglikelihood is extended to
\begin{align*}
l_1(\bs\beta,\sigma^2,\Sigma) = l_0(\bs\beta,\sigma^2,\Sigma) + \frac{K}{2} \sum_{m>1} \log |S_m| - \sum_{m>1} \sum_k \frac{(\widetilde{\bs\beta}_{mk} - \whbs\beta_m)^T S_m (\widetilde{\bs\beta}_{mk} - \whbs\beta_m)}{2\psi\widetilde{\sigma}_{mk}^2}.
\end{align*}
We can clearly see that the remote posterior samples provide additional information on $S_m$.

We still propose the same MLE as in \eqref{eqn:MLE} with $l_0$ replaced by $l_1$.
Note that only the E-step needs to be changed compared to the algorithm in Section \ref{sec:CEDAR0} as follows.
\begin{align*}
\bb E (S_m | \bs\beta,\sigma^2,\Sigma) = (n+K+1) (\Sigma^{-1} + A_m A_m^T)^{-1},
\end{align*}
where $A_m = \begin{bmatrix} \mb a_m & B_m/\sqrt{\psi} \end{bmatrix}$ with
$B_m =  \begin{bmatrix} (\widetilde{\bs\beta}_{m1} - \whbs\beta_m) / \widetilde{\sigma}_{m1} & \cdots & (\widetilde{\bs\beta}_{mK} - \whbs\beta_m) / \widetilde{\sigma}_{mK} \end{bmatrix}$.
This leads to Algorithm \ref{alg:CEDAR} with nonzero $K$ and $A_m$.
Note that, if $K > p$, the matrix $B_mB_m^T$ should be transferred to the central site, instead of $B_m$, to save communation cost.
In this case, the time complexity of the algorithm is $\mc O(Mp^3)$ per iteration.
On the other hand, if $K \le p$, we can use the Woodbury matrix identity, to replace the E-step with the following.
\begin{align*}
S_m \leftarrow (n+K+1) ( \Sigma - \Sigma A_m (I + A_m^T \Sigma A_m)^{-1} A_m^T \Sigma ),
\end{align*}
for which the time complexity reduces to $\mc O((K+1)Mp^2)$ per iteration.
Therefore, the general time complexity becomes $\mc O(\min(K+1,p)Mp^2)$ per iteration.

% Sparse Regularization

\subsection{Sparse regressions}

CEDAR can obtain sparse estimates by simply adding a penalty to the loglikelihood.
\begin{align*}
	l'(\bs\beta,\sigma^2,\Sigma) = l(\bs\beta,\sigma^2,\Sigma) - P(\bs\beta).
\end{align*}
The penalty $P$ can be any widely used penalty function such as Lasso \citep{Tibshirani1996}, SCAD \citep{Fan2001}, elastic net \citep{Zou2005}, MCP \citep{Zhang2010}, and so on.
This will only change the M-step for $\bs\beta$ in Algorithm \ref{alg:CEDAR}.
For example, if $L_1$ penalty $P(\bs\beta) = \lambda \|\bs\beta\|_1$ is employed, we can use the proximal gradient descent algorithm \citep{Beck2009} to update $\bs\beta$ as follows.
\begin{align*}
\bs\beta \leftarrow \prox_{\lambda \sigma^2,s} \left( \bs\beta - s\sum_m S_m (\bs\beta -  \whbs\beta_m) \right),
\end{align*}
where $\prox_{\lambda,s}$ is the proximal operator associated with the $L_1$ penalty defined as
\begin{align*}
\prox_{\lambda,s}(\mb x) = \argmin_\mb z \left( \frac{1}{2s} \|\mb z - \mb x \|_2^2 + \lambda \|\mb z\|_1 \right),
\end{align*}
and the step size $s$ is determined by the backtracking line search.

% Theory

\section{Theoretical Properties}
\label{sec:theory}

In this section, we investigate the differential privacy properties in the notion of \citet{Dwork2014} for the posterior samples.
Then, we present the asymptotic properties of CEDAR that will form the basis for inference.

\subsection{Differential Privacy of Posterior Samples}

Differential privacy is a probabilistic system for protecting private information in a dataset in sharing or processing the data.
A random procedure $f$, which takes a (deterministic) dataset $\mc D$ as input and generate a random output $f(\mc D)$, is said to be differentially private if the output cannot leak the information of any particular individual in the dataset.
To fix the idea, suppose there are two \emph{neighboring} datasets $\mc D_1$ and $\mc D_2$ where $\mc D_1 = (X,\mb y)$ has $n$ data points and $\mc D_2$ has one more data point $(\mb x,y)$ than $\mc D_1$, \emph{or vice versa}.
The property of differential privacy requires the distribution of $f(\mc D_1)$ and $f(\mc D_2)$ are similar enough that anyone cannot make an inference on the difference between $\mc D_1$ and $\mc D_2$, that is, the extra individual $(\mb x,y)$, based on the outcomes.
\citet{Dwork2014} formalizes this idea and define the $(\epsilon,\delta)$-DP property as follows.

\begin{definition}
    Let $\bs\beta = f(\mc D)$ be the outcome of a random mechanism which takes a dataset $\mc D$ as input.
    The mechanism is said to have the $(\epsilon,\delta)$-differential privacy property if, for any Borel set $B \subset \bb R^p$ and any neighboring datasets $\mc D_1$ and $\mc D_2$, we have
    \begin{align} \label{eqn:DPDef}
        P(\bs\beta \in B|\mc D = \mc D_1) \le e^\epsilon P(\bs\beta \in B|\mc D = \mc D_2) + \delta,
    \end{align}
\end{definition}

If \eqref{eqn:DPDef} holds with small $\epsilon$ and $\delta$, it means the two distributions $P(\bs\beta|\mc D_1)$ and $P(\bs\beta|\mc D_2)$ are nearly indistinguishable, and thus it is difficult to tell whether $\bs\beta$ was generated from $\mc D_1$ or $\mc D_2$.
Therefore, smaller $\epsilon$ and $\delta$ offer stronger guarantee of privacy protection.
In the literature, $\delta$ is typically set to $1/n$ where $n$ is the dataset size and $\epsilon<1$ is considered strong DP, $\epsilon<4$ is considered acceptable, and $\epsilon>8$ considered weak \citep{Abadi2016,Brendan2018}.

Noting that a set of posterior samples can be seen as an output of a random mechanism which takes a dataset as input, we shall show that it possesses the $(\epsilon,\delta)$-DP property, and therefore the remote posterior samples can carry the information of the remote dataset while protecting the individual patients' data of the remote site.
Let $S_1 = X^TX$ and $S_2 = S_1 + \mb x \mb x^T$ be the gram matrices for $\mc D_1$ and $\mc D_2$, respectively, and let $\xi_1 = \frac{1}{\psi\sigma^2} (\bs\beta_2 - \bs\beta_1 )^T S_1 (\bs\beta_2 - \bs\beta_1 )$, $\xi_2 = \frac{1}{\psi\sigma^2} (\bs\beta_2 - \bs\beta_1 )^T S_2 (\bs\beta_2 - \bs\beta_1 )$, and $c = \mb x^T S_1^{-1} \mb x$ where $\bs\beta_1$ and $\bs\beta_2$ are the OLS estimates for $\mc D_1$ and $\mc D_2$, respectively,

\begin{theorem} \label{thm:DP}
    Suppose $S_1$ is positive definite and $\bs\beta^{(1)},\dots,\bs\beta^{(K)}$ are independent posterior samples from $\pi_\psi(\bs\beta|\mc D)$ where $\mc D=\mc D_1$ or $\mc D=\mc D_2$.
    For any Borel sets $B_1,\dots,B_K \subset \bb R^p$, it follows that
    \begin{align*}
        \prod_{k=1}^K P(\bs\beta^{(k)} \in B_k|\mc D = \mc D_1) \le e^\epsilon \prod_{k=1}^K P(\bs\beta^{(k)} \in B_k|\mc D = \mc D_2) + \delta,
    \end{align*}
    if
    \begin{align} \label{eqn:eps}
        \epsilon > \epsilon_\delta \equiv -\frac{K}{2} \log (1+c) + \frac{Kc}{2} + \frac{K\xi_2}{2} + c \log(1/\delta) + c\sqrt{ K (1+2\lambda) \log (1/\delta)},
    \end{align}
    where $\lambda = (y-\mb x^T \bs\beta_1)^2/(\psi\sigma^2c)$,
    and it follows that
    \begin{align*}
        \prod_{k=1}^K P(\bs\beta^{(k)} \in B_k|\mc D = \mc D_2) \le e^\epsilon \prod_{k=1}^K P(\bs\beta^{(k)} \in B_k|\mc D = \mc D_1),
    \end{align*}
    if 
    \begin{align*}
        \epsilon > \frac{K}{2} \log (1+c) + \frac{K\xi_2}{2}.
    \end{align*}
\end{theorem}

Readers are referred to Supplementary Material for the proofs.
Theorem \ref{thm:DP} shows the mechanism of generating a posterior sample is $(\epsilon,\delta)$-differentially private and provides the lower bounds of $\epsilon$ for any $\delta$. 
The differential privacy of posterior samples has been discussed in \citet{Dimitrakakis2017} under the assumption that the Lipschitz constant for the loglikelihood is bounded or the prior is tight enough.
Note that their work is not applicable to our setting because the Lipschitz constant for the Gaussian loglikelihood with respect to data is unbounded and we consider the flat prior for $\bs\beta$.
Moreover, since the results in \citet{Dimitrakakis2017} do not discuss the magnitude of $\epsilon$ and $\delta$, it is unclear how strongly the data can be protected.
Theorem \ref{thm:Eeps} constructs the upper bounds (in probability) for $\epsilon_\delta$ in Theorem \ref{thm:DP}.

\begin{theorem} \label{thm:Eeps}
    Suppose $S_1$ is positive definitive.
    Assume the outcome variables $\mb y$ and $y$ are independent and follow the model assumption $\mb y \sim \mc N(X \bs\beta_0,\sigma^2 I)$ and $y \sim \mc N(\mb x^T \bs\beta_0,\sigma^2)$.
    Then, $\epsilon_\delta$ in Theorem \ref{thm:DP} satisfies
    \begin{align*}
        \bb E ( \epsilon_\delta | X,\mb x ) \le -\frac{K}{2} \log (1+c) + \frac{Kc}{2} + \frac{Kc}{2\psi} + c \log(1/\delta) + c\sqrt{ K (1+2(1+c)/(\psi c)) \log (1/\delta)}.
    \end{align*}
\end{theorem}

\begin{remark} \label{rem:epsdelta}
    Theorem \ref{thm:Eeps} implies that, if $Kc \rightarrow 0$, then $\epsilon_\delta$ in \eqref{eqn:eps} converges to 0 in probability, which can be shown by the Markov inequality. 
    In fact, it is typical to have $c \approx p/n$. Recalling $n$ is the size of the common dataset $\mc D_1$, the result confirms we have stronger privacy protection as $n\rightarrow \infty$ and weaker protection as $K\rightarrow\infty$.
    This clearly poses the tradeoff between the estimation efficiency and the privacy protection.
\end{remark}

Note that the bounds for $\epsilon_\delta$ is independent of $\sigma^2$ as the distributions of the quantities $\xi_2$ and $\lambda$ are independent of $\sigma^2$ under the assumption of Theorem \ref{thm:Eeps}.
Also note that Theorem \ref{thm:Eeps} provides a tighter bound of $\epsilon$ for the entire $K$ posterior samples than what the composition rule \citep{Dwork2014} can offer, which states we have $(\epsilon'=K\epsilon,\delta'=K\delta)$-DP property for the $K$ independent posterior samples when each posterior sample is $(\epsilon,\delta)$-DP.

\subsection{Asymptotic Properties and Inference}

CEDAR has appealing asymptotic properties which make statistical inference feasible.
Although the algorithm has been derived under the normality assumption for features and errors, the theoretical properties discussed in this section do not require such normality.
In addition to the model assumption \eqref{eqn:model}, we only impose very mild regularity conditions.
\begin{assumption} \label{ass:1}
$\sup_{1 \le m \le M} \left\| \frac{1}{n} X_m^T X_m - \Sigma_0^m \right\|_2 = o_p(1)$ as $n \rightarrow \infty$ where $\Sigma_0^m$ are symmetric and positive definite.
\end{assumption}

\begin{assumption} \label{ass:2}
For any $m \ge 1$, $\lambda_{\min} ( \Sigma_0^m ) > t > 0$ and 
$\lambda_{\max} ( \Sigma_0^m ) < T < \infty$.
\end{assumption}

\begin{assumption} \label{ass:3}
$\sup_n \sup_{1<m \le M} \left\| \frac{n}{K} B_m B_m^T - n(X_m^T X_m)^{-1} \right\|_2 = o_p(1)$ as $K \rightarrow \infty$.
\end{assumption}

\begin{assumption} \label{ass:4}
For any $n$, $\lim_{M \rightarrow \infty} \frac{1}{M} \sum_m \left( \frac{1}{n} X_m^T X_m - \Sigma_0^m \right) = 0$.
\end{assumption}

All assumptions can be satisfied under trivial situations.
Note that the feature vectors $\mb x_i^m$ can have heterogeneous variance $\Sigma_0^m$, and Assumptions \ref{ass:1} and \ref{ass:3} implicitly constrain $M$, the number of datasets.
Although our proofs will assume iid Gaussian errors with mean zero, it can easily be relaxed to more general error distributions such as sub-Gaussian with zero mean.
Let ($\whbs\beta$, $\wh\sigma^2$, $\wh\Sigma$, $\wh S_1$, \dots, $\wh S_M$) be the outcome of CEDAR.
%Note that $\wh \Sigma$ is the zero of the following function.
%\begin{align} \label{eqn:zero}
%   f(\Sigma) = \frac{1}{N} X_1^TX_1 + \frac{n+K+1}{N} \sum_{m>1}  (\Sigma^{-1} + \widehat{A}_m\widehat{A}_m^T)^{-1} - \Sigma.
%\end{align}

\begin{theorem} \label{thm:main}
Suppose Assumptions \ref{ass:1}-\ref{ass:3} hold.
Assume $n \rightarrow \infty$ and $K/n \rightarrow \gamma > 0$, and let $\Sigma_0$ be the zero of
\begin{align*}
   f_0(\Sigma) = \frac{1}{M} \Sigma_0^1 + \frac{1+\gamma}{M} \sum_{m>1}  (\Sigma^{-1} + \gamma (\Sigma_0^m)^{-1})^{-1} - \Sigma.
\end{align*}
Then, it follows that
\begin{enumerate}[(i)]
    \item $\widehat{\Sigma} - \Sigma_0 \rightarrow_p 0$,
    \item $\sqrt{N} \Sigma_0 \Sigma_*^{-1/2} ( \widehat{\bs\beta} - \bs\beta_0 ) \rightarrow_d \mc N(\mb 0, \sigma_0^2 I)$,
    \item $\widehat{\sigma}^2 \rightarrow_p  \sigma_0^2$,
\end{enumerate}
where
\begin{align*}
    \Sigma_* = \frac{1}{M} \Sigma_0^1 + \frac{(1+\gamma)^2}{M} \sum_{m>1} (\Sigma_0^{-1} + \gamma (\Sigma_0^m)^{-1})^{-1} (\Sigma_0^m)^{-1} (\Sigma_0^{-1} + \gamma (\Sigma_0^m)^{-1})^{-1},
\end{align*}
which can be consistently estimated by 
\begin{align*}
    \wh \Sigma_* = \frac{1}{N} \wh S_1 + \frac{1}{NK} \sum_{m>1} \wh S_m B_m B_m^T \wh S_m.
\end{align*}
\end{theorem}

All proofs can be found in Supplementary Material.
Note that the zero of $f_0(\Sigma)$ satisfies
$$
    \left( I + \gamma \sum_{m>1} \Sigma_0 ( \Sigma_0^m + \gamma \Sigma_0)^{-1} \right) \Sigma_0 = \Sigma_0^1 + \gamma \sum_{m>1} \Sigma_0 ( \Sigma_0^m + \gamma \Sigma_0)^{-1} \Sigma_0^m,
$$
which implies that $\Sigma_0$ is a weighted sum of $\Sigma_0^m$'s and thus bounded by Assumption \ref{ass:2}.
Similar to the result for differential privacy, the asymptotic behavior of CEDAR depends on the asymptotic ratio $\gamma$ of $K$ to $n$.
As $\gamma \rightarrow 0$, we have $\Sigma_0 \rightarrow \Sigma_0^1$ and $\Sigma_* \rightarrow \frac{1}{M} \sum_m \Sigma_0^1 (\Sigma_0^m)^{-1} \Sigma_0^1$.
On the other hand, as $\gamma \rightarrow \infty$, we have $\Sigma_0 \rightarrow \frac{1}{M} \sum_m \Sigma_0^m$ and $\Sigma_* \rightarrow \frac{1}{M} \sum_m \Sigma_0^m$.
These align with the results of Theorems \ref{thm:K0} and \ref{thm:finite}, respectively.

\begin{theorem} \label{thm:K0}
Suppose Assumptions \ref{ass:1}-\ref{ass:3} hold.
Assume $n \rightarrow \infty$ and $K/n \rightarrow 0$.
Then, it follows that
\begin{enumerate}[(i)]
    \item $\widehat{\Sigma} - \Sigma_0^1 \rightarrow_p 0$,
    \item $\sqrt{N} \Sigma_*^{-1/2} ( \widehat{\bs\beta} - \bs\beta_0 ) \rightarrow_d \mc N(\mb 0, \sigma_0^2 I)$,
    \item $\widehat{\sigma}^2 \rightarrow_p  \sigma_0^2$,
\end{enumerate}
where
\begin{align*}
    \Sigma_* = \frac{1}{M} \sum_m (\Sigma_0^m)^{-1},
\end{align*}
which can be, provided that $K \rightarrow \infty$, consistently estimated by 
\begin{align*}
    \wh \Sigma_* = \frac{n}{M} \wh S_1^{-1} + \frac{n}{MK} \sum_{m>1} B_m B_m^T.
\end{align*}
\end{theorem}

Theorem \ref{thm:K0} implies, when $K$ is relative smaller than $n$, the asymptotic behavior of $\whbs\beta$ is similar to that of the AVGM estimator and the variance estimator $\wh \Sigma$ consistently estimates $\Sigma_0^1$ only.
The asymptotic variance $\Sigma_*$ can still be estimated consistently if $K\rightarrow \infty$, but cannot be estimated otherwise due to the lack of information.
Corollary \ref{cor:homogeneous} considers the special case of Theorems \ref{thm:main} and \ref{thm:K0} where $\Sigma_0^m$ are homogeneous.

\begin{corollary} \label{cor:homogeneous}
Suppose Assumptions \ref{ass:1}-\ref{ass:3} hold.
Assume $n \rightarrow \infty$ and $K/n \rightarrow \gamma \ge 0$.
Further assume that $\Sigma_0^m$ are homogeneous, that is, $\Sigma_0^1=\cdots=\Sigma_0^M = \Sigma_0$.
Then, it follows that
\begin{enumerate}[(i)]
    \item $\widehat{\Sigma} - \Sigma_0 \rightarrow_p 0$,
    \item $\sqrt{N} \Sigma_0^{1/2} ( \widehat{\bs\beta} - \bs\beta_0 ) \rightarrow_d \mc N(\mb 0, \sigma_0^2 I)$,
    \item $\widehat{\sigma}^2 \rightarrow_p  \sigma_0^2$.
\end{enumerate}
\end{corollary}

Distributed learning is particularly useful when each site has a limited number of data but we have access to many remote datasets.
Thus, $n\rightarrow\infty$ may not be a feasible assumption from a practical viewpoint, and it becomes an important question what happens when we have a diverging number of datasets ($M \rightarrow \infty$) while $n$ is fixed.
The next result confirms that CEDAR can still be efficient.

\begin{theorem} \label{thm:finite}
Suppose Assumptions \ref{ass:2}-\ref{ass:4} hold.
Assume $M \rightarrow \infty$ and $K \rightarrow \infty$ while $n$ is fixed, and let $\Sigma_0 = \frac{1}{M} \sum_m \Sigma_0^m$.
Then, it follows that
\begin{enumerate}[(i)]
    \item $\widehat{\Sigma} - \Sigma_0 \rightarrow_p 0$,
    \item $\sqrt{N} \Sigma_0^{1/2} ( \widehat{\bs\beta} - \bs\beta_0 ) \rightarrow_d \mc N(\mb 0, \sigma_0^2 I)$,
    \item $\widehat{\sigma}^2 \rightarrow_p  \sigma_0^2$.
\end{enumerate}
\end{theorem}

\begin{remark} \label{rem:n}
    When $\Sigma_0^m$ are homogeneous, Theorem \ref{thm:finite} still holds with $\Sigma_0 \equiv \Sigma_0^m$.
    Again, it is important that Theorem \ref{thm:finite} holds without having $n \rightarrow \infty$, while AVGM and CSL \citep{Jordan2019} require $n \rightarrow \infty$ to achieve the optimal asymptotic variance. To see it, for example, \eqref{eqn:AVGMVar} implies
    \begin{align*}
        N \Var \left(\whbs\beta^A \right) \rightarrow \frac{n}{n-p-1} \sigma_0^2 \Sigma_0^{-1},
    \end{align*}
    as $M \rightarrow \infty$ when $n$ is fixed.
    Theorem \ref{thm:finite} can hold because CEDAR can incorporate information from remote sites via posterior samples.
    Of course, we cannot have arbitrarily large number of remote posterior samples in practice if the data privacy issue is in place, as also noted in Remark \ref{rem:epsdelta}.
    However, even a small $K$ brings substantial improvements as we will see in Section \ref{sec:simulation}.
\end{remark}

Based on above results, when $\Sigma_0^m$ are homogeneous, CEDAR suggests the following Wald test statistic for the null hypothesis $H_0: \beta_{0j} = b_{0j}$ against $H_a: \beta_{0j} \neq b_{0j}$.
Under $H_0$, we have
\begin{align} \label{eqn:Wald}
	W_j = \frac{\sqrt{N}(\wh\beta_j - b_{0j})}{\wh\sigma \sqrt{(\wh\Sigma^{-1})_{jj}}} \sim \mc N(0,1),
\end{align}
suggesting rejecting $H_0$ if $|W_j| > z_{1-\frac{\alpha}{2}}$, where $\alpha$ is the significance level and $z_q$ is the $q$-quantile of the standard normal distribution.
Note that CEDAR can perform other tests such as the test of a contrast without additional communication.
We can also construct the $100(1-\alpha)\%$ confidence interval for $\beta_j$ as follows.
\begin{align*}
\left( \wh\beta_j - z_{1-\frac{\alpha}{2}} \wh\sigma \sqrt{(\wh\Sigma^{-1})_{jj}/N}, \wh\beta_j + z_{1-\frac{\alpha}{2}}  \wh\sigma \sqrt{(\wh\Sigma^{-1})_{jj}/N} \right). 
\end{align*}

% Simulation

\section{Simulation}
\label{sec:simulation}

We compare CEDAR with existing approaches in three aspects; estimating the regression coefficients, sparse regression, and inference.
The competitors include the average mixture approaches (\citet{Zhang2013} for estimating regression coefficients and \citet{Battey2018} for hypothesis tests), CSL \citep{Jordan2019}, and OPT based on euqation \eqref{eqn:OPT}.
We consider two versions of the CSL approaches; CSL$_1$ and CSL$_A$.
CSL$_1$ takes the local OLS estimator at the central site as its initial estimator $\overline{\bs\beta}$, and CSL$_A$ uses AVGM as the initial estimator, which demands an extra communication with remote sites.
We also include multiple versions of CEDAR. CEDAR$_K$ uses $K$ remote posterior samples where $K=0,4,16$.
For each dimension $p$ considered ($p=4,32$), we generate 100 datasets as follows.
A half of the predictors (randomly chosen) are generated from Gaussian $\mc N(0, 1)$, a quarter of the predictors are generated from Uniform $\mc U(-\sqrt{3},\sqrt{3})$, and the remaining quarter are generated from the Laplace distribution $\mc L(\sqrt{2})$.
The regression coefficients and the response variable are generated from
\begin{align*}
\beta_{0j} \sim \begin{cases} \mc U(0,1), & 1 \le j \le p/4,\\
0, & p/4 < j \le p, \end{cases} \qquad y_i^m|\mb x_i^m,\bs\beta_0 \sim \mc N(\bs\beta_0^T \mb x_i^m ,\sigma_0^2).
\end{align*}
Note that $\Sigma_0^m = I$ for all $m$ and $\sigma_0^2=1$ was used.
The average performance measures over the 100 simulated datasets are reported for each method.

We first examined the privacy protection levels of the posterior samples for CEDAR$_4$ and CEDAR$_{16}$.
Table \ref{tbl:DP} shows the minimum $\epsilon$ values satisfying \eqref{eqn:DPDef}, which are obtained by Monte Carlo simulation, are reported for the CEDAR versions considered where $c=p/n$ and $\delta=1/n$. They are strongly differently private except for when $c \approx 1$.

\begin{table}[!ht]
    \centering
    \begin{tabular}{cc||cccccc}
    \hline
    $p$ & $K$ & & $c=1$ & $c=1/2$ &  $c=1/4$ &  $c=1/8$ &  $c=1/16$ \\
    \hline
    \hline
    \multirow{2}{*}{4} & 4 & & 0.57 & 0.41 & 0.25 & 0.14 & 0.09\\
    & 16 & & 3.38 & 1.76 & 0.90 & 0.47 & 0.26\\
    \hline
    \multirow{2}{*}{16} & 4 & & 3.28 & 1.73 & 0.91 & 0.48 & 0.27\\
    & 16 & & 7.88 & 3.86 & 1.93 & 1.00 & 0.54\\
    \hline
    \end{tabular}
    \vskip 0.2in
    \caption{Differential privacy levels of the posterior samples in simulation. The (minimum) $\epsilon$ values are reported where $c=p/n$, $\delta=1/n$, and $\psi=100$.}
    \label{tbl:DP}
\end{table}

\begin{figure}[!ht]
\begin{center}
\includegraphics[width=0.9\textwidth]{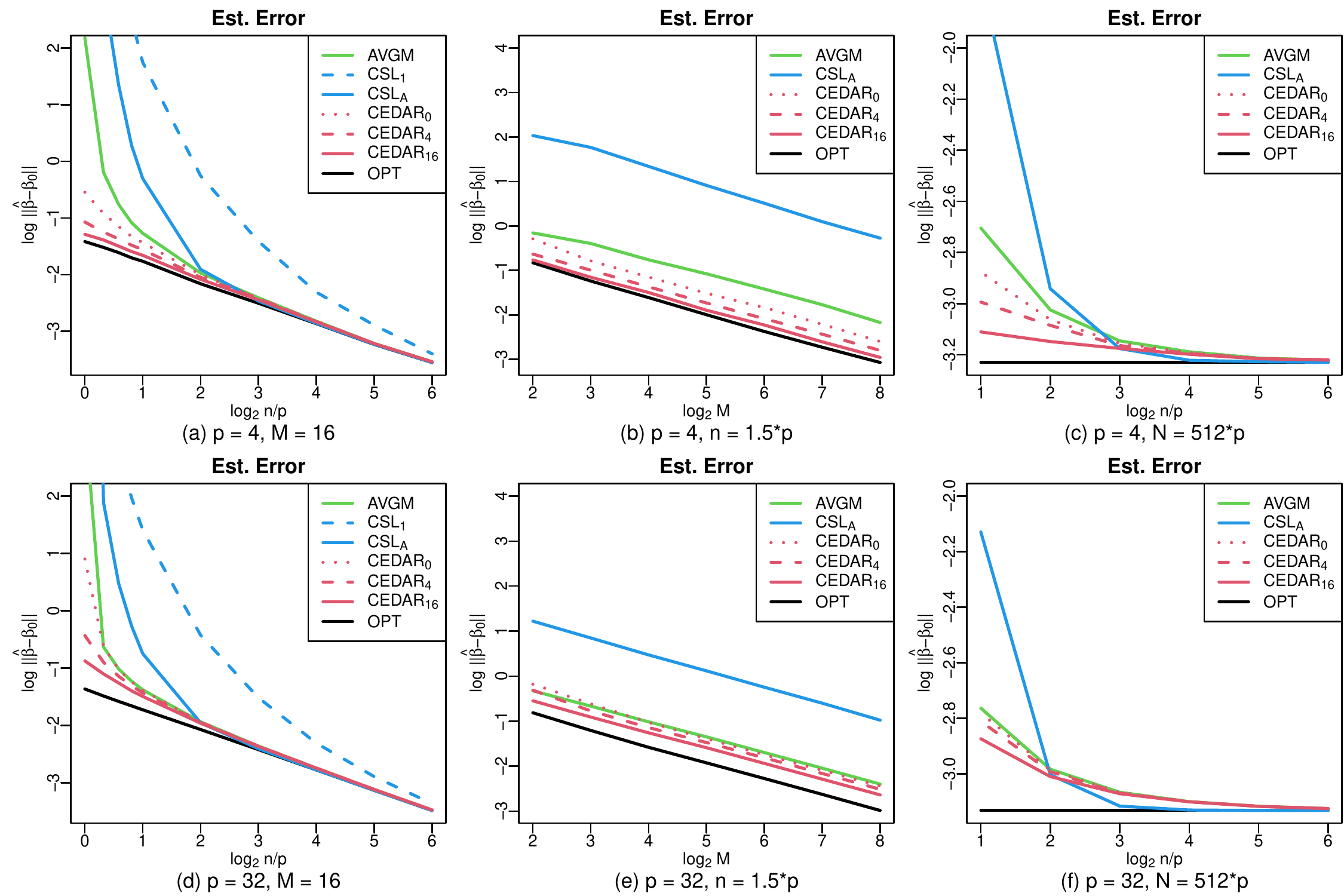}
%\vskip -0.1in
\caption{Simulation results in estimating the regression coefficients. Average $L_2$ distance of the estimates from the true regression coefficients are plotted.}
\label{fig:est}
\end{center}
\end{figure}

Figure \ref{fig:est}(a) and \ref{fig:est}(d) show how the estimation error (log average $L_2$ norm) of the regression coefficients changes as the sample size varies, while the number of sites is fixed.
All methods work well when $n \ge 4p$ except for CSL$_1$.
(For clear presentation, we have removed CSL$_1$ from all the remaining plots.)
However, CEDARs outperform when $n < 4p$ and also  perform better as $K$ increases.
Particularly when $p$ is small, even CEDAR$_0$, which does not use any remote posterior samples, outperforms AVGM and CSLs.
Figure \ref{fig:est}(b) and \ref{fig:est}(e) show how the estimation error changes as the number of remote sites increases, while the sample size is fixed.
All lines decreases roughly in parallel except for CSL$_1$. 
This implies no method can attain the optimal asymptotic variance if $n$ is fixed, confirming Remark \ref{rem:n}.
Only CEDAR can get close to OPT as $K$ increases.
In Figure \ref{fig:est}(c) and \ref{fig:est}(f), we fix the total sample size $N$, while the number of sites and the sample size vary.
It is confirmed again that CEDARs outperform AVGM and CSLs when $n$ is small.
We also notice that CSL$_A$ has an outstanding performance when $n \ge 8p$, which however requires an additional communication.

\begin{figure}[!ht]
\begin{center}
\includegraphics[width=0.9\textwidth]{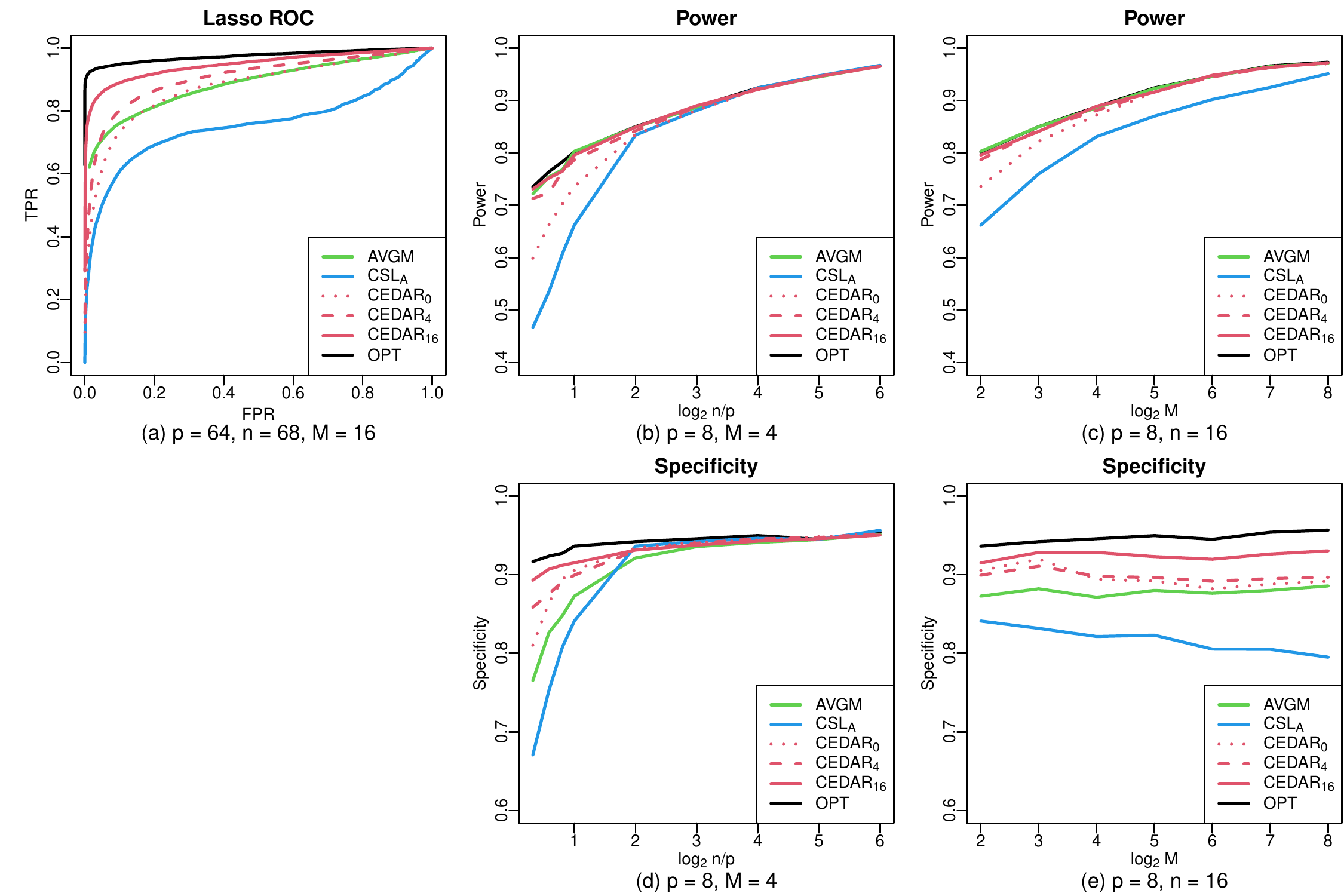}
\vskip -0.1in
\caption{(a) ROC curve for variable selection in sparse regressions; (b)-(e) Average specificity and power in hypothesis tests.}
\label{fig:inference}
\end{center}
\end{figure}

To compare the performance of variable selection through penalized regressions, we evaluate the lasso type of regularized estimates for all methods, except for AVGM becuase it is not amenable to the $L_1$ regularizer. Instead, we apply the hard-thresholding on the AVGM estimates.
OPT is equivalent to the regular lasso estimator using the full pooled data.
Figure \ref{fig:inference}(a) shows the receiver operating characteristic (ROC) curve for variable selection. 
We note that CEDAR with $K>0$ shows outstanding variable selection performance.

To compare performance in statistical inference, we conduct the Wald test.
Similar to \eqref{eqn:AVGMWald}, we use the approach of \citet{Battey2018} as an inference version for AVGM.
The CSL methods use the equation (13) in \citet{Jordan2019} for estimating the Fisher information, which requires no additional communication.
OPT uses the standard Wald test statistic for the full pooled data.
CEDAR uses \eqref{eqn:Wald}.
We test the null hypothesis $H_0: \beta_{0j} = 0$ against $H_a: \beta_{0j} > 0$, for all $j$ \emph{individually}.
The significance level was set to $\alpha=0.05$.
The power to detect nonzero coefficients is estimated from the tests on the $p/4$ nonzero coefficients out of the 100 simulated datasets, and the specificity is calculated from the tests on the $3p/4$ zero coefficients.

Figure \ref{fig:inference}(b) and \ref{fig:inference}(d) show how the power and the specificity change as the sample size $n$ changes, while the number of sites is fixed.
The power increases as the sample size increases for all methods, but CEDAR is the only one other than OPT that does well for both power and specificity.
Figure \ref{fig:inference}(c) and \ref{fig:inference}(e) show how the power and the specificity change as the number of sites varies, while the site sample size is fixed.
All methods suffer a discount in specificity, but only CEDAR, as $K$ increases, shows accurate specificity.
Overall, CEDAR shows better performance in terms of both power and specificity than other distributed analysis methods.

\section{Application}
\label{sec:data}

We illustrate how practically useful CEDAR can be when applied to the analysis of real distributed data.
We use the Georgia Coverdell Acute Stroke Registry (GCASR) data that cover nearly 80\% of acute stroke admissions in Georgia, USA from 2005 to 2013.
As analyzed in \citet{Deng2016}, we fit the linear model that predicts the arrival-to-CT time of patients using 14 selected features; 'NIHStrkS', 'EMSNote', 'LipTotal', 'Age', 'Gender', 'RaceAA', 'RaceW', 'HlthInsM', 'Day', 'NPO', 'MedHisST', 'MedHisTI', 'MedHisVP", and 'MedHisFHSTK'.
There are 42 hospitals which have more than 15 patients, with all patients data that have missing values removed. 
To highlight the difference of each method, we simulate 3 different settings which vary with the combination of site sample sizes. In the first setting, we only include the hospitals with at most 80 patients, which results in 626 patients from 13 hospitals.
In the second setting, we include hospitals with at most 500 patients, which results in 4272 patients from 31 hospitals.
In the third setting, we include all hospitals, which results in 13300 patients from 42 hospitals.
In all settings, we choose the hospital with the median number of patients as the central site.

\begin{figure}[!ht]
\begin{center}
\includegraphics[width=0.9\textwidth]{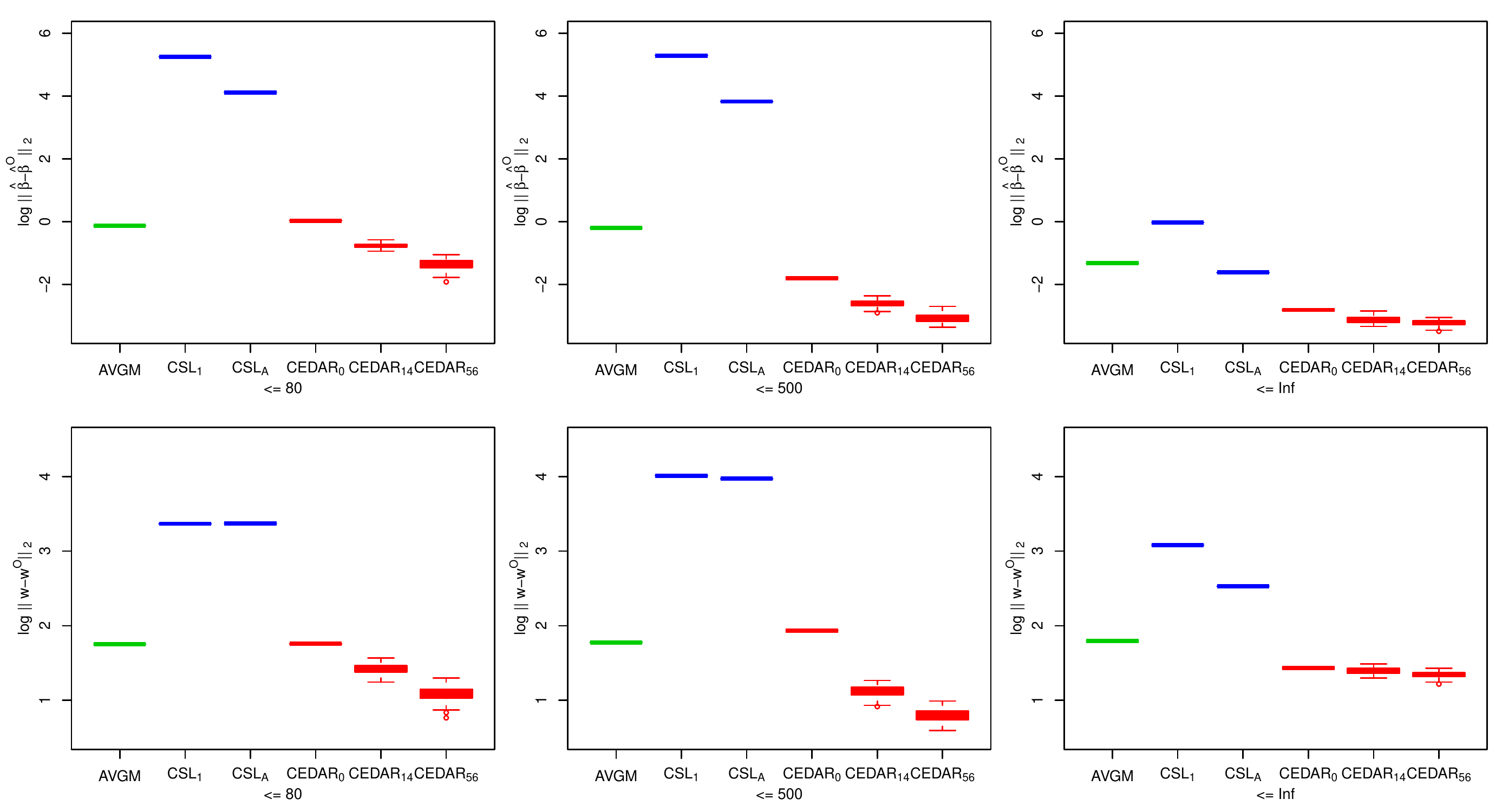}
\vskip -0.1in
\caption{Boxplots of distance between estimated coefficient and the OLS coefficient (upper panels) and boxplots of distance between Wald test statistic vectors (lower panels) for three different settings. The left panels are obtained when the hospitals with at most 80 patients are included. The middle panels are obtained when the hospitals with at most 500 patients are included. The right panels are obtained when all hospitals are included.}
\label{fig:application}
\end{center}
\end{figure}

As we do not have the gold standard, each method is compared to the full data OLS estimator (OPT).
Noting that CEDAR is a stochastic procedure due to the randomness of posterior samples,  CEDAR with a nonzero number of posterior samples is performed on 100 replicates, each of which has a different set of posterior samples.
Figure \ref{fig:application} shows the $L_2$ distances from the estimates obtained by each method to the optimal estimate $\whbs\beta^O$ (upper panels), and shows the $L_2$ distances from the Wald test statistics vector obtained by each method to the optimal Wald test statistics vector $\whmb w^O$.
Again, AVGM for Wald test uses the method in \citet{Battey2018}.

Overall, the results are commensurate with those from the simulation study.
In terms of estimation, all methods show improvements as we include more hospitals.
Meanwhile, CEDAR shows consistently good performance while CSL shows the most dramatic changes as the sample size cap and the sample size of the central site increase.
In terms of inference, it is not necessarily the case that all methods improves even if we include more hospitals.
It is perhaps due to the heterogeneity across datasets, which accumulates as we include more hospitals and poses challenges to the estimation of the Fisher information.
Particularly, the CSL approach suffers greater challenges since it only uses the information from the central site to estimate the Fisher information.

\section{Discussion}
\label{sec:discussion}

We have proposed CEDAR, an MLE-based method for distributed regression models.
CEDAR provides enhanced statistical efficiency by aggregating the information from external data through the remote MLEs and the remote posterior samples.
This framework enables the proper statistical inference that is highly important for the analysis of EHRs data and can easily accommodate the sparse regressions.
CEDAR is communication efficient, has desirable theoretical properties for statistical inference and privacy protection, and is shown to outperform the existing communication-efficient distributed learning methods in numerical studies, particularly when the site sample sizes are small.

One potential limitation of CEDAR compared to the competing methods is its computational costs.
If the remote posterior sample size $K$ is small, CEDAR can be as computationally efficient as others.
But, the costs can be much higher if both $p$ and $K$ large.
Nevertheless, our experiments suggest that even a moderate $K$ can make a significant improvement.

We focused on the linear regression in this article because the idea can be delivered in the clearest way.
The algorithm derivations are exact with no approximation and the theoretical properties of CEDAR inherit the unbiasedness.
But, the principle of CEDAR is not limited to the linear regression.
A future research direction is to build a more general framework encompassing more complex models including the generalized linear models.

\section*{Acknowledgements}

This work is partly supported by NIH grant R01GM124111.
The content is solely the responsibility of the authors and does not necessarily represent the official views of the National Institutes of Health.\vspace*{-8pt}

\bibliographystyle{apalike}
\bibliography{CEDAR}  %%% Uncomment this line and comment out the ``thebibliography'' section below to use the external .bib file (using bibtex) .





% \end{thebibliography}

\end{document}